\documentclass[twocolumn,usenatbib]{mn2e}

\usepackage{natbib}
\usepackage{amsbsy}
\usepackage{amssymb}
\usepackage{amsmath}

\usepackage{graphicx}

\newcommand{\be}{\begin{equation}}
\newcommand{\ee}{\end{equation}}
\newcommand{\bd}{\begin{displaymath}}
\newcommand{\ed}{\end{displaymath}}
\newcommand{\bear}{\begin{eqnarray}}
\newcommand{\eear}{\end{eqnarray}}

\newcommand{\rn}{{\rm n}}
\newcommand{\rp}{{\rm p}}

\newcommand{\cE}{{\cal E}}

\newcommand{\rK}{{\rm K}}

\newcommand{\cR}{{\cal R}}

\newcommand{\cN}{{\cal N}}

\begin{document}

\title[r-mode damping and saturation] {A new mechanism for saturating unstable r-modes in neutron stars}

\author[Haskell, Glampedakis \& Andersson]{B. Haskell$^{1,2}$, K. Glampedakis$^{3,4}$ \& N. Andersson$^5$ \\
  \\
  $^1$ Max-Planck-Institut f\"{u}r Gravitationsphysik, Albert-Einstein-Institute, Am M\"{u}hlenberg 1, Potsdam, D-14776, Germany \\
  $^2$ School of Physics, The University of Melbourne, Melbourne, Victoria 3010, Australia\\
  $^3$ Departamento de F\'isica, Universidad de Murcia, Murcia, E-30100, Spain \\
  $^4$ Theoretical Astrophysics, University of T\"{u}bingen, Auf der Morgenstelle 10, T\"{u}bingen, D-72076, Germany \\
  $^5$ Mathematical Sciences and STAG Research Centre, University of Southampton, Southampton SO17 1BJ, UK}

\maketitle

\begin{abstract}
We consider a new mechanism for  damping the oscillations of a mature neutron star. The new dissipation channel arises if superfluid vortices are forced to cut through superconducting fluxtubes. This mechanism is interesting because the oscillation modes need to exceed a critical amplitude in order for it to operate. Once it acts the effect is very strong (and nonlinear) leading to efficient damping. The upshot of this is that modes are unlikely to ever evolve far beyond the critical amplitude.  We consider the effect of this new dissipation channel on the r-modes, that may be driven unstable by the emission of gravitational waves. Our estimates show that the fluxtube cutting leads to a saturation threshold for the instability that can be smaller than that of other proposed mechanisms. This suggests that the idea may be of direct astrophysical relevance.  \end{abstract}


\section{CONTEXT}
\label{sec:intro}

Neutron stars represent a hands-off laboratory for physics under extreme conditions, and may ultimately provide a complement to information gleaned from particle colliders like the LHC. While such terrestrial experiments probe hot plasmas at relatively low densities, the core of a neutron star requires an understanding of the cold dense part of the QCD phase diagram \citep{alford08}. To gain access to this information we need to accurately model how a realistic neutron star interior connects to its exterior and affects observable features. 
	
A commonly considered example involves the cooling of the star. Which processes lead to the star cooling down and how does heat flow from the interior to the surface? By matching  models of possible scenarios to X-ray data for isolated neutron stars, we may be able to constrain the theory. An excellent recent example of this is provided by the observed real-time cooling of the remnant in Cassiopeia A, which has provided the first true constraint on the superfluid transition temperature for the star's core \citep{shternin,CasA}. 

Another aspect of the problem relates to the dynamics of the star's complex core. A neutron star undergoes a number of changes as it evolves and provided that these are dramatic enough, various stellar oscillation modes may be excited. These could, in turn, affect the emission pattern of the star (either in X-ray or radio) provided that the interior fluid motion has a significant effect on the star's magnetosphere. This has led to the development of neutron star astero-seismology, where the aim is to use future observations to probe the star's interior in the same way that helio-seismologists have successfully constrained the interior physics on the Sun. 

A breakthrough in this area came with the observations of quasi periodic oscillations in the X-ray tails of large magnetar flares \citep{sw}. Early, relatively naive, models suggested that the observed oscillations could be identified with various elastic oscillation modes of the star's crust \citep{piro05,samuelsson07}. More recent work has attempted, not yet completely successfully, to account for the anticipated strong magnetic field effects \citep{colaiuda12,gabler13}. This is a very difficult problem, but there has been clear progress in the last few years. 

Since neutron stars are distant, one would expect their oscillations to be excited to detectable amplitudes only under exceptional circumstances. Such events would be rare, like the magnetar flares. However, there is an exception to this rule. Modes of oscillation may become unstable at various instances during the star's life. Provided an unstable mode is allowed to grow large enough, such instabilities may lead to a detectable signal and may also have an indirect effect on the star's evolution (say of the spin). A number of possible instabilities have been discussed in the literature. As far as mature neutron stars are concerned, the most promising ideas involve the Coriolis driven r-modes, which somewhat counter-intuitively may become unstable due to the gravitational waves they emit \citep{andersson98, friedman98}. This has stimulated a large body of work on the nature of the r-modes, the gravitational wave signal they would be associated with and the physics that may affect the growth of the instability. A number of possible damping and saturation mechanisms have been suggested over the last 15 years or so \citep{arras03,nayyar06, BTW07,BTW09,MF, HYP, CFL, luciano1, bildsten00, glampedakis06, gusakov13}. Nevertheless, the conclusions from state-of-the-art modelling remain relatively unaffected. The r-mode instability is likely to set a spin-threshold for neutron stars. This is an important observation since the fastest observed radio pulsars and accreting neutron stars spin well below the theoretical break-up limit \citep{chak03,patruno10}. A mechanism is required to explain this, and the r-mode instability appears to fit the bill. Furthermore a recent analysis of the problem has shown that the theoretical predictions for the r-mode instability window for a 'minimal' neutron star model, which does not include superfluity or the appearance of exotic particles (such as hyperons or deconfined quarks) in the core, is not consistent with current X-ray observations of Low Mass X-ray Binaries (LMXBs) \citep{ho11, HDH}. There is, therefore, a clear need to include additional effects in our modelling, such as superfluity and superconductivity in the core of the star.

This paper introduces a new mechanism to the r-mode scenario. The argument involves the star's core and builds on the fact that there is likely to be a region where superfluid neutron vortices co-exist with superconducting protons. As has been argued in different contexts, such a region may have decisive impact on the star's dynamics. Due to the interaction between superfluid vortices and magnetic fluxtubes, any changes in the star's vorticity (the bulk rotation or the fluid motion associated with an oscillation mode) may be coupled to the magnetic field. This suggests  two scenarios. In the first, the vortices become pinned to the, more plentiful, fluxtubes. In the second scenario, the vortices can cut through the flux tubes, but at a cost. This latter process is expected to be highly dissipative.  It is this possibility that we explore in this paper.

\section{Brief summary}

The fact that superfluid dynamics is damped by a mutual friction arising from the interaction between quantised vortices and other components in the mixture (typically, phonons in laboratory studies of He$_4$ and electrons in a neutron star core) is well established. The main idea dates back to work by Hall and Vinen in 1955 \citep{HV}. They introduced a linear friction between superfluid (Helium) vortices and the ``normal'' component (represented by phonons). Balancing this force to the Magnus force that would drive the vortices to move along with the superfluid condensate in the absence of friction, they deduced the functional form for the force,  $ \mathbf{F}_{\rm mf}$ in the following, that couples the two ``fluid'' components in the system; the superfluid condensate and the normal component. 

In the standard picture, the vortex  friction, $\mathbf{f}_\mathrm{D}$,  is taken to be \emph{linear} in the relative velocity $\mathbf{u}$ between the vortices and the normal fluid;
\be
\mathbf{f}_\mathrm{D}=  \rho_\rn \kappa \cR \mathbf{u}
\label{drag}\ee
where  $\rho_\rn$ and $\kappa$ are, respectively, the  density of the superfluid and the quantum of circulation associated with each vortex. The dimensionless
friction coefficient, $\cR$, is assumed to be velocity-independent. The force balance that controls  the motion of individual vortices
leads to a linear algebraic relation $\mathbf{u} = \mathbf u(\mathbf{w})$, where $\mathbf w$ is the relative velocity between the condensate and the normal fluid. Inverting this relation one finds that the relative fluid flow is damped 
according to 
\be
\partial_t \mathbf{w} + \{ ... \} = -\frac{1}{x_\rp \rho_\rn}  \mathbf{F}_{\rm mf}
\label{releq}
\ee
where $ \mathbf{F}_{\rm mf}$ is obtained from $\mathbf f_\mathrm{D}$ by using the inferred $\mathbf u (\mathbf w) $ relation and combining the effect for an array of vortices. The  brackets in \eqref{releq}  represent fluid terms that are not relevant to this discussion and $x_\rp = \rho_\rp/\rho$ is the
normal fluid  fraction ($\rho = \rho_\rp + \rho_\rn $ is the total density).


In the case of superfluid neutron star dynamics, one can show that a similar relation applies provided that $\rho_\rn$ and $\rho_\rp$ are taken to be the neutron and proton densities. Hence, a relation like (\ref{releq}) will affect the relative motion associated with any global oscillation mode. This means that we can  extract a characteristic mutual friction dissipation timescale in terms of the mode energy
$E_{\rm mode}$ (obtained as a volume integral of the inviscid velocity field) and the rate of work $\dot{E}_{\rm mf}$ done by $\mathbf{F}_{\rm mf}$. Provided the damping rate is slow compared to the dynamics of the mode, the timescale is well approximated by: 
\be
\tau_{\rm mf} = \frac{2 E_{\rm mode}}{|\dot{E}_{\rm mf}|}
\label{time1}
\ee
This timescale can be very short if the mode under consideration has a significant counter-moving component. Detailed work has shown that this is the case for  the fundamental f-mode, and as a result the gravitational-wave driven instability of those modes is severely suppressed in a superfluid star \citep{LM95,fmode}. The conclusions for  the Coriolis restored r-mode is different. These modes are affected by mutual friction to a much lesser extent, essentially because they are mainly horizontal \citep{lee03, MF, passamonti09}. 

The mechanism  we consider in this paper is subtly different from the Hall-Vinen model in that the friction force turns out to be \emph{non-linear} in the relative 
flow $\mathbf{u}$. This means that the inferred mode damping, still expressed in terms of an $\cR$ coefficient, will be velocity dependent. 
Whenever this is the case, the problem has interesting new aspects. Most importantly, the equation for the relative motion \eqref{releq} becomes nonlinear, which means that the mutual friction may be able to prevent a given oscillation mode  from growing beyond some threshold amplitude. That is, in addition to damping the mode, the mutual friction may lead to the saturation of an instability. 

In the following section we outline the derivation of the new friction force. The steps involved essentially repeat the analysis of \citet{link03}. Having done this, we will discuss the implications for the r-mode instability. Readers that are mainly interested in the astrophysical results (or may already be familiar with the fluxtube cutting mechanism) can proceed straight to section 4.


\section{The new friction mechanism}
\label{sec:friction}

\subsection{Vortex-fluxtube pinning}
\label{sec:pin}

The interaction between superfluid neutron vortices and superconducting proton fluxtubes in the outer core of a neutron star is thought to be key to the evolution of the system, possibly linking  changes in spin to the evolution of the magnetic field. 
An important ingredient in this problem is the energy cost associated with superfluid vortices, which are expected to be magnetised due to the entrainment effect \citep{als88}, cutting through superconducting fluxtubes. As a rough estimate one may consider the 
energy associated with superposition of a neutron vortex and a proton fluxtube. This leads to what we will refer to as the 
pinning energy, $f_\mathrm{pin}$, acting on each moving vortex. Ignoring geometrical factors related to direction dependence, 
the force per intersection is of the order \citep{ruderman}
 \be
F_\mathrm{int} \approx {E_\mathrm{int} \over \Lambda_*} = \Lambda_*^2 B_\rn B_\rp 
\ee
where the London penetration length $\Lambda_*$ (which is of the order of few tens of $fm$) represents the typical size of the overlap region, while $B_\rn$ and $B_\rp$ are the magnetic fields carried by individual vortices and fluxtubes, respectively. The force per unit length of a given vortex is then
\be
f_\mathrm{pin} \approx { F_\mathrm{int} \over d_\rp}
\ee
where
\be
d_\rp \approx \left( {B \over \phi_0} \right)^{1/2} \approx 3\times10^3 B_{12}^{-1/2} \ \mathrm{fm}
\ee
Here $B$ (and $B_{12}=B/10^{12}$G) is the macroscopic core magnetic field, $\phi_0$ is the quantum of magnetic flux and $d_\rp$ is the typical distance separating the fluxtubes. 

This estimate allows us to quantify how easy it is for a vortex to cut through the array of fluxtubes in a neutron star core.  A necessary condition is that vortices
do not pin onto the fluxtubes, which means that the Magnus force must exceed the pinning force. 
To make this quantitative, let us represent the vortex and fluxtube velocities by $\mathbf{u}_\rn$ and $\mathbf{u}_\rp$, respectively. Meanwhile, the macroscopic flows (that enter the averaged two-fluid hydrodynamics) are given by $\mathbf{v}_\rn$ (for the superfluid neutrons) and $\mathbf{v}_\rp$ (for the proton condensate). 
If a vortex is pinned to the fluxtubes, then we expect to have $\mathbf{u}_\rn =\mathbf{u}_\rp \approx \mathbf{v}_\rp$. Basically, it is natural to assume that the fluxtubes move with the proton condensate. This means that the velocity difference that enters into the Magnus force is approximated by 
$\mathbf{u}_\rn- \mathbf{v}_\rn \approx \mathbf{v}_\rp -\mathbf{v}_\rn \equiv \mathbf{w}$
Given this, we can obtain a \emph{minimum} velocity lag, $w_\mathrm{pin}$, between the neutron and proton fluids {\em below which} vortex pinning is likely to take place~\citep{link03}
\be
w_{\rm pin} \approx \frac{f_{\rm pin}}{\rho_\rn \kappa} 
\approx  1.5 \times 10^4\, \, B_{12}^{1/2} \,~\mbox{cm}/\mbox{s}
\label{dwpin}
\ee
We note here that in this expression (and the ones hereafter) only the dependence with respect to the magnetic field is shown while the fluid density 
has been set to $\rho=10^{14}$ g/cm$^3$, a value representative of a neutron star outer core. 

The estimate (\ref{dwpin}) will be of central importance later. The key point is that, as long as the relative velocity $w$ between the two fluids
is below $w_\mathrm{pin}$, the vortices will not be able to move relative to the fluxtubes. Hence, the damping mechanism that we will now discuss will not act.


\subsection{Kelvin-wave damping}
\label{sec:cut}

Once the pinning can no longer balance the Magnus force and the vortices start moving, they must cut through the fluxtube array to keep going. 
This may be a highly dissipative process due to the excitation of Kelvin waves along the vortex. This point was first argued by \citet{epstein92} 
for vortices moving through the lattice of nuclei in the star's crust, and later adapted by \citet{link03} to the conditions in the core that we discuss here.
In an effective theory, the waves on the vortex  can be treated as particles, ``kelvons'', with effective mass $\mu$ and energy $E_k = \hbar^2 k^2/2\mu$, where $k$ is the associated wavenumber.
If we let
\be
\mathbf{u} = \mathbf{u}_\rp- \mathbf{u}_\rn
\ee
be the relative vortex-fluxtube velocity, then the interaction at each intersection lasts a time interval
$t_{\rm int} \sim \Lambda_*/u$. The kelvon energy can be estimated by using this characteristic timescale in the standard formula for an  oscillator; $E_k\approx \hbar/t_\mathrm{int}$ (ultimately originating from the uncertainty principle). This then leads to the  characteristic wavenumber
\be
k \approx \left ( \frac{2\mu}{\hbar \Lambda_*} u \right )^{1/2} \equiv \frac{1}{\Lambda_*} \left ( \frac{u}{v_\Lambda} \right )^{1/2}
\ee
Given that the characteristic velocity
\be
v_\Lambda = \hbar / 2 \mu \Lambda_* \approx 10^9 \ \mathrm{cm/s}
\ee
we should typically have $k \Lambda_* \ll 1$ in the case of neutron star dynamics. A more sophisticated analysis, leading to the same final estimate, can be found in~\citet{link03}.

In order to calculate a dissipation rate we need the kelvons produced at different intersections of the same vortex to add incoherently. This requires $k d_\rp \gg 1$, which in turn leads to a lower limit for the relative vortex-fluxtube velocity;
\be
u_{\rm low} \approx  6.5 \times 10^5 \, B_{12}\,~ \mbox{cm}/\mbox{s}
\ee
In order for the mechanism we discuss to operate efficiently we need $u\gg u_\mathrm{low}$. Note that $u_\mathrm{low}>w_\mathrm{pin}$ when $B\gtrsim10^8$~G or so. The estimates we present are thus still  consistent for the case of LMXBs, as long as the {\it internal} magnetic field is not much stronger than the inferred {\it exterior} dipolar magnetic field strength (which is inferred to be $\approx 10^8$~G). If we want to consider significantly stronger magnetic fields we would need to first understand the behaviour at velocities in the range between  $w_\mathrm{pin}$ and  $u_\mathrm{low}$  better. 

The energy released at each vortex/fluxtube intersection was determined 
by \citet{link03}. The result is
\be
\Delta E = { 2 \over \pi} { F_\mathrm{int}^2 \over \rho_\rn \kappa} \left( v_\Lambda u \right)^{-1/2}
\ee
This suggests that the energy loss rate (per unit volume) is 
\be
\dot{\cE}_\mathrm{cut} =  {\cN_\rn u \over d_\rp^2} \Delta E  
\ee
where $\cN_\rn$ is the number of vortices per unit area. Alternatively, we can use the fact that (ignoring entrainment, which only affects the estimate by a factor of order unity \citep{na06})
 $ 2\Omega_\rn \approx \cN_\rn \kappa$, to get
\be
\dot{\cE}_\mathrm{cut} =    {4 \Omega_\rn \over \pi \rho_\rn \kappa^2} f_\mathrm{pin}^2 \left( {u \over v_\Lambda} \right)^{1/2}
\label{dEcut} \ee

We can relate the rate (\ref{dEcut}) to the work done by a drag force (exerted on a unit length vortex segment)
of the general form
\be
\mathbf{f}_{\rm D} = \rho_\rn \kappa \cR  \mathbf{u}
\label{fv}
\ee 
with a velocity-dependent coefficient $\cR = \cR(u)$. Then from
\be
\dot{\cE}_{\rm cut} = \mathbf{f}_{\rm D} \cdot \mathbf{u}
\ee
we infer that
\be
\cR = \cR_0 \left ( \frac{v_\Lambda} {u}\right )^{3/2}
\label{Rdrag}
\ee
with
\be
\cR_0 =  \frac{2}{\pi} \left( \frac{f_{\rm pin}}{\rho_\rn \kappa v_\Lambda} \right )^2 
\quad \to \quad \cR_0 \approx 1.4 \times 10^{-10} \,  B_{12}
\label{R0drag}
\ee

Finally, we can construct
a hydrodynamical mutual friction force density exerted on the neutron fluid by averaging the force (\ref{fv}) over the vortex array.
This leads to
\be
 \mathbf{F}_{\rm mf} = \cN_\rn \mathbf{f}_{\rm D} \quad \to \quad  \mathbf{F}_{\rm mf} = 2\Omega_\rn \rho_\rn \cR_0  
\left ( \frac{v_\Lambda} {u}\right )^{3/2} \mathbf{u}
\label{Fmf}
\ee

The key observation here is that, as soon as the vortices start to move relative to the fluxtubes they are likely to be prevented by a very strong friction. This is obvious since $w_\mathrm{pin}$ and $u_\mathrm{low}$ are both going to be much smaller than $v_\Lambda$. The lower the relative velocity, the stronger this damping is.  In practice, this means that the vortices are unlikely to be able to keep moving and the system will be driven back towards pinning.


\section{r-mode damping and saturation}
\label{sec:rmode}

In the previous section we outlined the argument that leads to vortices cutting though fluxtubes being a highly dissipative process. This argument is not original, but we believe this is the first time that the  discussion has been framed in the context of a mutual friction force. The final result \eqref{Fmf} allows us to consider the mechanism in a range of relevant contexts. For example, once the dissipation due to vortex-fluxtube cutting is expressed as a mutual friction force we can adapt it for the two-fluid hydrodynamics model used to model neutron star oscillations and instabilities. As an illustration of this analysis, let us try to estimate what the effect on the gravitational-wave driven r-mode instability may be.

In order to make use of the deduced mutual friction force in a problem involving the standard two-fluid model, we first of all need to replace the dependence on the relative velocity, $\mathbf u$, between vortices and fluxtubes with the relative velocity, $\mathbf w$, between the two macroscopic fluid components. The standard approach to this, pioneered by Hall and Vinen more than half a century ago~\citep{HV}, is to first balance the vortex force \eqref{Fmf} by the Magnus force that acts on the vortices and the invert the relation to get an expression for $\mathbf u = \mathbf u (\mathbf w)$. The steps involved are straightforward in the case where the friction coefficient $\mathcal R$ is constant.  When $\mathcal R$ is velocity dependent the analysis becomes slightly more involved and one should in principle consider the full problem, including relative flows in the background. However, in the present case we can bypass this problem by making a couple of (potentially debatable) assumptions.

First of all, on dynamical timescales the fluxtubes can be assumed to move with the protons, which means that $\mathbf{u}_\rp \approx \mathbf{v}_\rp$.
It is not quite so easy  to justify a similar relation between the neutron fluid and vortex velocities. To make progress, we nevertheless 
\emph{assume} that $ \mathbf{u}_\rn \approx   \mathbf{v}_\rn$. This  would
be  true for free vortices and it might be a reasonable approximation in the case of vortices moving at high speed through the
fluxtube array. This is, in fact, the approximation underpinning the model in Section~\ref{sec:friction} so it make sense to make this approximation here as well.
With these  assumptions we simply have $\mathbf{u} = \mathbf{w}$.

Now, from detailed two-fluid calculations \citep{MF} we know that  unstable r-modes have a particular relative velocity contribution. 
In general this contribution is position dependent, due to the density dependence of the superfluid pairing gaps. 
In order to keep things simple, we will nevertheless assume that this contribution is proportional to the average velocity perturbation, $\mathbf v$. This leads to 
\be
w = \lambda v \quad \to \quad w \approx \lambda \alpha \Omega R
\ee
where $\alpha$ is the usual (dimensionless) r-mode amplitude (e.g.~\citet{owen98}). We know from actual mode calculations that the counter-moving contribution enters at higher order in the slow-rotation expansion such that 
\be
\lambda =  \lambda_0 \left ( \frac{\Omega}{\Omega_\rK} \right)^2
\ee
where $\Omega_\rK$ is the break-up frequency and $\lambda_0 $ is taken to be a spin-independent factor. We have  ignored the radial dependence of the r-mode's velocity field, 
$ v \sim (r/R)^2$, which  should be reasonable as long as the main damping effect originates in the outer core of the star where this factor is of order unity.
This is, of course, a simplification but in this first proof-of-principle discussion we prefer to proceed analytically rather than turn to numerical solutions. Given this attitude, we feel that this is a natural simplification to make.


If the mode has large enough amplitude to force vortices through fluxtubes, then  $w \gtrsim w_{\rm pin}$ which means that
\be
\cR \lesssim  2.5 \times 10^{-3}\,  B_{12}^{1/4}
\ee
It is worth noting that the deduced upper limit is about a factor $\sim 10-100$ larger than the (velocity-independent) drag coefficient
associated with the standard mutual friction mechanism; scattering of electrons by vortices \citep{als88,na06}.

The corresponding damping timescale can be estimated in the usual way (see e.g. \citet{nakk01}) by making use of  (\ref{time1}).
This argument involves the r-mode energy 
\be
E_{\rm mode} \approx \frac{1}{2} \alpha^2 \Omega^2 M R^2 \tilde{J}
\qquad \mbox{where} \qquad 
\tilde{J} = \frac{1}{M R^4}\int_0^R \rho r^6 dr 
\ee
which leads to $\tilde{J} = 0.016$ for an $n=1$ polytrope~\citep{owen98}. 
The mutual friction damping rate is given by 
\be
\dot{E}_{\rm mf} = \int  \dot{\cE}_{\rm cut} dV
\ee
and our estimates lead to
\be
\dot{E}_{\rm mf} \approx \frac{4 \Omega}{\pi \kappa^2} \int  \frac{f_{\rm pin}^2}{\rho}
\left ( \frac{w}{ v_\Lambda} \right  )^{1/2} dV
\ee
That is,
\be
\dot{E}_{\rm mf} \approx  \frac{8 \Omega^{5/2}}{\pi f_{\rm K}} \frac{f_{\rm pin}^2}{\kappa^2}
\left ( \frac{\alpha \lambda_0}{R v_\Lambda} \right )^{1/2}  \int^R_{R_{\rm in}}  \frac{r^3}{\rho} dr
\ee
We have assumed that fluxtube cutting takes place in the outer part of the stellar core, in the region  $R_{\rm in} < r \lesssim R$ (where the coexistence
of a neutron superfluid and a proton superconductor is likely) and that $\lambda_0$ and $\rho$ are approximately uniform. 

Through these arguments we obtain an order of magnitude estimate for the mutual friction damping timescale 
\be
\tau_{\rm mf} \approx 6 \times 10^{10} \lambda_0^{-1/2} \alpha^{3/2} \nu_{500}^{-1/2} B_8^{-1}\,~ \mbox{s}
\label{time2}
\ee
where $\nu_{500} = \nu/500$~Hz is the scaled spin-frequency of the star ($\nu=\Omega/2\pi$). 

If we want to consider the relevance of the proposed mechanism for various astrophysical scenarios, then we need to provide an estimate for   $\lambda_0$. This will require a more detailed numerical calculation for realistic superfluid parameters etcetera. However, we can use previous mode-calculations to get an idea of the likely range of values for this parameter. Extracting an averaged value from the  r-mode study by \citet{MF} (assuming their pinning limit) we find that $\lambda_0$ ought to lie in the range
\be
\langle \lambda_0 \rangle \approx 0.1 - 1
\label{lam0}
\ee
both for strong and weak superfluidity models. This result is obviously not very precise, but it will allow us to assess whether the new damping mechanism is strong enough to warrant a more detailed investigation.


In considering possible astrophysical scenarios it is important to appreciate that the features of the new mechanism are rather different from the standard mutual friction. 
Most importantly, the dissipation due to fluxtube cutting is a non-linear process that  {\em saturates} but does not completely
suppress an unstable mode. This mutual friction mechanism does not operate as soon as $w$ is driven down to $ w_{\rm pin}$
when vortices can repin  to the fluxtubes. Hence, one would expect an unstable mode to evolve in such a way that its amplitude saturates around this level. 
This provides a rough estimate of the r-mode amplitude of such systems
\be
w \approx w_{\rm pin} \quad \to \quad 
\alpha_{\rm pin} \approx 10^{-6}\, \left ( \frac{\lambda_0}{0.1} \right )^{-1} \nu_{500}^{-3} B_8^{1/2} 
\label{alpha_min}
\ee
We can use this threshold amplitude to rewrite the mutual friction timescale (\ref{time2}) in a more transparent form
\be
\tau_{\rm mf} \approx 190 \, \left ( \frac{\lambda_0}{0.1} \right )^{-2} \left ( \frac{\alpha}{\alpha_{\rm pin}} \right)^{3/2}  
\nu_{500}^{-5} B_8^{-1/4}\,~ \mbox{s}
\ee
This timescale is (at least) about an order of magnitude shorter than the mode's growth timescale, assuming a $n=1$ polytropic star \citep{nakk01}.
It is therefore likely that the scenario outlined above works: once the mode amplitude exceeds $\alpha_{\rm pin}$ the unpinned vortex array is driven 
through the fluxtubes and the ensuing friction quickly damps out the mode, effectively suppressing it back to $\alpha_{\rm pin}$.


\section{Astrophysics: application to accreting systems}
\label{sec:astro}

An obvious astrophysical setting where the fluxtube cutting scenario may apply is in fast spinning accreting neutron stars in Low-Mass X-ray Binaries. From previous considerations of the  r-mode instability in this context \citep{bbr}, we know that the mode amplitude required to achieve torque balance is  
\be
\alpha_{\rm acc} \approx 1.3 \times 10^{-7} \left ( \frac{L_{\rm acc}}{10^{35}\,\mbox{erg}/\mbox{s}} \right )^{1/2}
\nu_{500}^{-7/2}
\ee
Balancing the two mechanisms, as would be appropriate if the fluxtube cutting allows the r-mode to grow to the precise amplitude required to prevent further spin-up in an accreting system, we have
\be
\frac{\alpha_{\rm pin}}{\alpha_{\rm acc}} \approx 8 \, \left ( \frac{\lambda_0}{0.1} \right )^{-1}
 B^{1/2}_8 \left ( \frac{L_{\rm acc}}{10^{35}\,\mbox{erg}/\mbox{s}} \right )^{-1/2}
\nu_{500}^{1/2}
\label{alpha_ratio}
\ee
In order for the new mechanism to play a role in explaining the observed population, one would expect to have  $\alpha_\mathrm{pin}/\alpha_\mathrm{acc}\approx 1$ for the fastest spinning systems.

As an example, let us consider 4U1608-522 which spins at 620~Hz and for which the averaged accretion luminosity is $5\times 10^{36}$~erg/s. If the range we have suggested for $\lambda_0$  is reliable, then we find that the proposed scenario would work provided the \emph{interior} magnetic field in this system is:
\be
B\approx (0.9-3)\times 10^8\ \mathrm{G}
\ee
This is in the range of the expected \emph{surface} fields for these mature systems. Moreover, it is natural to assume that the interior field (which may initially be much stronger than the externally visible field) of an old neutron star would be of the same order of magnitude as that in the exterior. The main point here is that our rough estimates lead to a results that appears consistent with both observations and our understanding of these systems. This makes it plausible that the new mechanism does, indeed, have a role to play in the r-mode scenario. At the very least, it warrants a more detailed investigation.

It is also worth noting an alternative strategy. We could take $\lambda_0$ as a ``free parameter'', which would make sense given our general ignorance of the conditions in the outer core of a neutron star. This parameter could then be constrained by observations relating to the magnetic field of fast spinning accreting neutron stars. As an example of this strategy, let us consider the data for IGR J00291+5934 (taking the observational constraints from~\citet{patruno10}). In this case, we have a spin frequency of 600~Hz, a luminosity of $6\times 10^{36}$~erg/s and a suggested external field of $B\approx 2\times 10^8$~G from the spin down rate in quiescence. From \eqref{alpha_ratio} we find that the accretion torque could be balanced by the fluxtube cutting mechanism as long as
\be
\lambda_0 \approx 0.16
\ee
comfortably inside the range suggested by the mode calculations. Again, this example suggests that the new mechanism should be relevant.

Finally, it is interesting to compare the mode amplitude $\alpha_{\rm pin}$ (essentially the saturation amplitude associated with
fluxtube cutting) against previous results on r-mode saturation due to non-linear couplings with other inertial modes. In general,
different saturation mechanisms would be competing with each other, and the one leading to the smallest mode amplitude would be 
physically the most relevant.   

The first incarnation of r-mode saturation by non-linear mode-couplings is the model of~\citet{arras03}; that work predicts
a maximum r-mode amplitude 
\be
\alpha_{\rm A} \approx 1.4 \times 10^{-3}\nu_{500}^{5/2} 
\label{arras}
\ee
This result has been refined by the more recent calculations of ~\citet{BTW07,BTW09}, resulting in a saturation amplitude
$\alpha_{\rm sat} \approx 0.1 \alpha_{\rm A}$ (for simplicity we retain the spin dependence of eqn.~(\ref{arras}) but we note
that the behaviour of the~\citet{BTW07,BTW09} saturation amplitude shows a rather complicated behaviour as a function
of time). 

Comparing this more recent mode-coupling saturation amplitude with our $\alpha_{\rm pin}$,
\be
\frac{\alpha_{\rm pin}}{\alpha_{\rm sat}} \approx 10^{-2} \left ( \frac{\lambda_0}{0.1} \right )^{-1} B_8^{1/2} \nu_{500}^{-11/2} 
\ee
This suggests that the fluxtube cutting mechanism is competitive as an r-mode saturation mechanism, being at least as efficient as
mode-coupling. This result supports our earlier claim that a more detailed investigation of the physics
of fluxtube-vortex interaction as a source of friction for the r-mode instability is needed.



\section{Concluding discussion}
\label{sec:discussion}

To summarize our results, we have formulated a new type of vortex mutual friction force, based on
the dissipative cutting of fluxtubes by fast moving vortices, and have studied its impact on the
r-mode instability in superfluid neutron stars. The non-linear dependence of this force with respect to the relative vortex-fluxtube
velocity leads to a rapid damping of the r-mode above a threshold amplitude at which the vortex array is forced to unpin 
from the fluxtubes. Effectively, this fluxtube-cutting friction provides a natural saturation mechanism for the r-mode
instability.  

We have highlighted the fact that our results may have important implications for the physics of accreting neutron stars in LMXBs.  
We have shown that the saturation amplitude due to fluxtube cutting (represented by $\alpha_{\rm pin}$, see eqn.~(\ref{alpha_min}))
could be smaller than the maximum amplitude set by non-linear couplings between the r-mode and
other inertial modes. Remarkably, this same amplitude could also be comparable to that required for balancing the accretion spin-up torque.
In practice this means that the saturation amplitude we calculate is such that it may allow for gravitational wave emission to be setting the spin equilibrium period for some of the hotter, faster, systems (which are in fact the best candidates for gravitational wave detection).  
However, for slower neutron stars in LMXBs, our amplitude may be small enough to never allow the mode to grow to the point where gravitational wave emission would influence the spin evolution (or, indeed, thermal evolution) of the system. This would allow a system to 'live' inside the standard r-mode instability window, without the need of additional damping mechanisms to explain the observations of \citet{HDH} and \citet{mah13}.

A more sophisticated treatment of the problem is of course necessary to accurately predict the relative strength of gravitational wave, accretion and electromagnetic spin down torques. This is of key importance for gravitational wave detection, given that recent analysis have shown that the dynamics of many sources is probably dictated by electromagnetic and accretion torques, with only a few systems likely to be interesting targets for next generation gravitational wave detectors \citep{hp11, patruno12, mah13}


There are aspects of the fluxtube-cutting friction that have not been discussed in any detail here. For instance,
an important issue is the fact that, as any
other frictional force, the mechanism discussed here should provide an additional source of heating in the stellar interior.  
However, calculating the rate of heating is a difficult task because the quasi-stationary state of the system is likely to be that
of pinning. This `pinning regime' may not actually translate to physically immobilised vortices. The system's finite temperature may
drive vortex creep with $u \sim w_{\rm pin}$. Unfortunately, in this velocity regime our analysis breaks down, making it 
impossible to make any prediction about dissipation and heating. We can, nevertheless, obtain an upper limit for the heating rate
by using the mode damping rate of the cutting regime. By then balancing the energy dissipation rate in (\ref{dEcut}) with the energy carried away by neutrino neutrino emission due to Cooper paring, $\dot{E}_{cp}=1.5\times 10^{31} T_8^8 \mbox{erg/s}$, with $T_8$ the temperature in units of $10^8$ K, we obtain core temperatures of order $10^8$ K. This temperature is consistent with the observed surface temperatures of LMXBs, especially for the faster systems which are also likely to be the most interesting for gravitational wave emission \citep{HDH, mah13}.

A more detailed understanding of  vortex-fluxtube interactions over the entire range of the expected velocities would represent a key advance in this area, relevant for  many aspects of neutron star dynamics. The mechanism we have discussed may not only be crucial for our understanding of the nonlinear development of the r-mode instability; it could also impact on models of pulsar glitches \citep{link13,pierreSUP} and  the combined magneto-rotational evolution of neutron stars \citep{ruderman, superconl, supercon}. To make further progress we need to sharpen our computational tools and develop models that account for the mesoscopic vortex-fluxtube interactions while, at the same time, track the macroscopic fluid dynamics. This is a challenging problem but the estimates we have presented provide clear motivation for future efforts.


\section*{Acknowledgments}
KG is supported by the Ram\'{o}n y Cajal Programme of the Spanish Ministerio de Ciencia e Innovaci\'{o}n 
and by the German Science Foundation (DFG) via SFB/TR7. 
BH is supported by the Australian Research Council (ARC) via a Discovery Early Career Award (DECRA).
NA is supported by STFC in the UK.


\end{document}